\documentclass[pre,preprint,showpacs,eqsecnum, amssymb,amsmath]{revtex4-1}
\usepackage{graphicx}

\begin{document}

\title{Bistability induced by two cross-correlated Gaussian white noises}

\author{A. N. Vitrenko}

\email{a.vitrenko@oeph.sumdu.edu.ua}

\affiliation{Department of General and Theoretical Physics, Sumy State University, 2, Rimskiy-Korsakov Street, 40007 Sumy, Ukraine}

\begin{abstract}
A prototype model of a stochastic one-variable system with a linear restoring force driven by two cross-correlated multiplicative and additive Gaussian white noises was considered earlier [S.~I.~Denisov \textit{et al}., Phys.~Rev.~E \textbf{68}, 046132 (2003)]. The multiplicative factor  was assumed to be quadratic in the vicinity of a stable equilibrium point. It was determined that a negative cross-correlation can induce nonequilibrium transitions. In this paper, we investigate this model in more detail and calculate explicit expressions of the stationary probability density. We construct a phase diagram and show that both additive and multiplicative noises can also generate bimodal probability distributions of the state variable in the presence of anti-correlation. We find the order parameter and determine that the additive noise has a disordering effect and the multiplicative noise has an ordering effect. We explain the mechanism of this bistability and specify its key ingredients.
\end{abstract}


\maketitle

\section{Introduction}\label{sec:intr}
Abstract (prototype, "toy") models of nonlinear systems play an important role in different fields of natural science and engineering such as physics, chemistry, biology, electronics and nanotechnology. On the one hand, they are simple enough to treat analytically in comparison with realistic models. On the other, they clearly demonstrate remarkable features of the system's behavior. Furthermore, abstract models are usually not related to any particular experimental system, but they can help us answer the question "What is possible?" \cite{Pelesko.2007}. As a few outstanding examples one may mention the Brusselator model for chemical oscillations \cite{Nicolis.1977}, a series of models proposed by R\"{o}ssler for chaotic dynamics \cite{Gurel.1983}, and coupled map lattices for a variety of phenomena in complex systems \cite{Kaneko.2001}.

A macroscopic nonequilibrium system is inherently open and subject to fluctuations of its environment or so-called "external noise"\cite{Moss.1989}. The former can often be described by several variables, and an equivalent noise-driven dynamical system exhibits stochastic behavior \cite{SchimanskyGeier.1997, Anishchenko.2007}. It is generally difficult to determine its statistical properties. Moreover, noise in nonlinear systems can lead to not only disorder that is obviously, but also to temporal or spatial order that is counterintuitive. In the latter case it plays constructive role and induces numerous phenomena that are impossible in the underlying deterministic dynamics. They are called noise-induced \cite{Wio.2003,Ridolfi.2011} and their examples include stochastic resonance \cite{Gammaitoni.1998, Wellens.2004, Rajasekar.2016}, Brownian ratchets \cite{Reimann.2002,Cubero.2016}, noise-induced transitions \cite{Horsthemke.1984}, noise-induced phase transitions \cite{GarciaOjalvo.1999, Sagues.2007}, etc.

The classical theory of noise-induced transitions was developed by Horsthemke and Lefever \cite{Horsthemke.1984}. It is based on the assumptions, considerably simplifying its construction. They are as follows: (i) the macroscopic system is spatially homogeneous (it is called zero-dimensional); (ii) the system can be described by one state variable; (iii) the external noise is stationary, commonly Gaussian and white. Consequently, the state variable is a Markovian diffusion process governed by Langevin and Fokker-Planck equations \cite{Coffey.2012, Risken.1989}, and its stationary probability density (SPD) can be obtained exactly. A noise-induced transition occurs, if the SPD of the state variable is changed qualitatively as the noise intensity exceeds a critical value. The genetic model \cite{Arnold.1978} and Hongler's model \cite{Hongler.1979} are relevant examples demonstrating this phenomenon. Their SPDs are changed in shape from unimodal to bimodal and two preferential states appear. The corresponding probabilistic potential becomes bistable, whereas the deterministic one is monostable. 

Subsequently, noise-induced transitions are discussed in many contexts. It is known that they arise from the multiplicative nature of the external noise. Additive white noise does not modify qualitatively the SPD of one-variable systems, but it can give rise to transitions in nonlinear two-variable oscillators \cite{SchimanskyGeier.1985, Landa.2000, Semenov.2016}. Small-size systems generate intrinsic multiplicative noise and discreteness-induced transitions can emerge \cite{Togashi.2001, Biancalani.2014, Houchmandzadeh.2015, Saito.2015, Saito.2016}. They have two different mechanisms, one of them is closely related to the classical theory. Noise-induced transitions in zero-dimensional systems are not phase ones in the thermodynamic sense. Noise-induced phase transitions are found in spatially extended systems \cite{GarciaOjalvo.1999, Sagues.2007} in which temporal components either do not exhibit \cite{VandenBroeck.1994, VandenBroeck.1997} or exhibit \cite{Ibanes.2001} noise-induced bistability.

Systems can be driven by two external cross-correlated noises \cite[and references therein]{Mendez.2014}. The influence of the cross-correlation on the stationary behavior of zero-dimensional one-variable systems is briefly studied in Ref.~\cite{Denisov.2003}. A simple model with a monostable deterministic potential driven by additive and multiplicative Gaussian white noises is considered. It is shown that the SPD of the state variable may be changed from unimodal to bimodal as the strength of a negative correlation between noises is varied. Besides, each noise does not induce transitions individually. In this paper, we provide a more detailed investigation of this model. We are interested how it is related to other known models that demonstrate noise-induced transitions. To this end, we explicitly calculate the SPD of the state variable and find its asymptotics. It is unclear why a negative cross-correlation can lead to a qualitative change of the stationary behavior of the system and what role noises play in this phenomenon. We expect that both additive and multiplicative noises can induce bistability when a negative cross-correlation is present. To confirm this, we plot a phase diagram and find the order parameter for transitions that corresponds to the extreme points of the SPD. We analyze a statistically equivalent one-noise model to understand the mechanism of bistability induced by two cross-correlated Gaussian white noises and to specify its key ingredients.

The paper is organized as follows. We present the model in Sec.~\ref{sec:model}. In Sec.~\ref{sec:SPD}, we obtain explicitly the SPD of the state variable and find its asymptotics. In Sec.~\ref{sec:PhD_OP}, we construct the phase diagram, find the order parameter, and discuss the mechanism of bistability. We summarize our results in Sec.~\ref{sec:Concl}.

\section{Model}\label{sec:model}

The considered dimensionless Langevin equation has the form \cite{Denisov.2003} (we slightly change the notation and rescale the equation in order to reduce the number of parameters)
\begin{equation}
	\label{eq:SDE1}
	\dot{x} = -x + \sigma_{1} x^{2} (1+x^{2})^{-1} \xi_{1}(t) + \sigma_{2} \xi_{2}(t),
\end{equation}
where the dot denote the derivative with respect to $t$, $x(t)$ is the state variable, $\xi_{1}(t)$ and $\xi_{2}(t)$ are cross-correlated multiplicative and additive Gaussian white noises. They have zero mean, intensities $\sigma_{1}^{2}/2$ and $\sigma_{2}^{2}/2$, respectively, and correlation functions
\begin{gather}
	\langle \xi_{1}(t)\xi_{1}(t') \rangle = \delta(t-t'), \quad
	\langle \xi_{2}(t)\xi_{2}(t') \rangle = \delta(t-t'), \nonumber\\
	\langle \xi_{1}(t)\xi_{2}(t') \rangle = r\delta(t-t'),\label{eq:NCF}
\end{gather}
where the brackets $\langle\cdot\rangle$ denote a statistical average, $\delta(t)$ is the Dirac delta function, and  $r$ is the coefficient of the correlation between the noises, $|r|\leqslant 1$. The Stratonovich interpretation of the Langevin equation (\ref{eq:SDE1}) is used. We assume natural boundary conditions for the state variable $x(t)$, it takes values both positive and negative including zero.

The linear restoring force $f(x)=-x$ corresponds to the parabolic potential $V(x)=x^{2}/2$ that is monostable with an equilibrium point at $x=0$. The multiplicative factor has the specific form $g(x)=x^{2}/ (1+x^{2})$. It is quadratic for small values of $|x|$ \cite{comment.1} and constant for large ones. The latter is means that $g(x)$ does not grow faster than linearly with $x$ and the stochastic process $x(t)$ does not explode \cite{Horsthemke.1984}. Though the appropriate multiplicative factor arises here from the mathematical considerations, it finds applications, for example, in models of ecological outbreak dynamics \cite{Sharma.2015}, genetic regulatory systems \cite{Liu.2009, Sharma.2016}, and tumor-immune system interactions \cite{Bose.2009, dOnofrio.2013}. The Langevin equation (\ref{eq:SDE1}) can be interpreted as describing the one-dimensional motion of an overdamped particle in the potential $V(x)$ subject to multiplicative $\xi_{1}(t)$ and additive $\xi_{2}(t)$ noises \cite{Gitterman.1999, Gitterman.2015}. This simple model demonstrates the interesting phenomenon of transitions induced by two cross-correlated noises when the SPD of $x(t)$ becomes bimodal though the associated deterministic dynamics ($\sigma_{1}=0$ and $\sigma_{2}=0$) is monostable. If $\sigma_{1}=0$ then $x(t)$ is the well-known Ornstein-Uhlenbeck process and its SPD is bell-shaped.

The Fokker-Planck equation for the probability density $P(x,t)$ of $x(t)$, statistically equivalent to Eqs.~(\ref{eq:SDE1}) and (\ref{eq:NCF}), is given by
\begin{equation}\label{eq:FPE}
	\frac{\partial }{\partial t} P(x,t)=-\frac{\partial }{\partial x} A(x)P(x,t)+\frac{\partial ^{2} }{\partial x^{2} } B(x)P(x,t), 
\end{equation}
where the drift $A(x)$ and diffusion $2B(x)$ coefficients have the form
\begin{eqnarray}
	A(x)&=&-x+\frac{1}{2} B'(x),\nonumber\\
	B(x)&=&\frac{1}{2}\sigma_{1}^{2} \left\{ [x^{2} (1+x^{2} )^{-1} +\alpha ]^{2} +\beta ^{2} \right\}. \label{eq:B(x)}
\end{eqnarray}
Here the prime denotes the derivative with respect to $x$, and, for convenience, we introduce new parameters as follows:
\begin{equation}
	\label{eq:alpha_beta)}
	\alpha = r\nu, \quad \beta =\nu\sqrt{1-r^2},\quad \nu =\frac{\sigma_{2}}{\sigma_{1}}.
\end{equation}

The diffusion coefficient (\ref{eq:B(x)}) is everywhere positive, if $-1<r\leqslant 1$ or $r=-1$ and $\sigma_{2}\geqslant\sigma_{1}$. In these cases, the state variable is unbounded, $x(t)\in(-\infty,\infty)$. If $r=-1$ and $\sigma_{2}<\sigma_{1}$, the diffusion coefficient is equal to zero at the points $\pm\sqrt{\sigma_{2}/(\sigma_{1}-\sigma_{2})}$. They are natural boundaries  and the domain of the state variable is restricted to the interval
\begin{equation}
	\label{eq:domen)}
	\left(-\sqrt{\sigma_{2}/(\sigma_{1}-\sigma_{2})},\sqrt{\sigma_{2}/(\sigma_{1}-\sigma_{2})}\right).
\end{equation}
We assume above that $\sigma_{2}\neq 0$. Otherwise $B(x)=0$ at $x=0$ where the restoring force vanishes as well. It is easy to see that this boundary point is absorbing (it can also be supported by an analysis of the boundary conditions \cite[pp.~104-107]{Horsthemke.1984}). Indeed, for large $|x|$ ($|x|\gg 1$), the noise $\xi_{1}(t)$ is additive and the restoring force is sufficiently strong to drive the system toward the steady state $x=0$. For small $|x|$ ($|x|\ll 1$), the multiplicative factor $g(x)$ is proportional to $x^{2}$ and the restoring force dominates again the stochastic force. If the particle reaches the absorbing point $x=0$, it remains there forever. The SPD of $x(t)$ becomes the Dirac delta function. The additive noise $\xi_{2}(t)$ destroys this boundary, keeping the system away from the steady state.

\section{Stationary probability density}\label{sec:SPD}

We write the stationary solution of the Fokker-Planck equation (\ref{eq:FPE}) as
\begin{equation}
	\label{eq:SPD0}
	\mathcal{P}(x)=\mathcal{N}[B(x)]^{-1/2}\exp[-\Phi (x)],
\end{equation}
where $\mathcal{N}$ is the normalizing factor and $\Phi (x)$ is the modified
potential. Up to a constant, the latter is given by 
\begin{equation}
	\label{eq:PF0}
	\Phi (x) =\sigma_{1}^{-2}\int\limits_{}^{x^{2} } \frac{(y^{2}+2y+1)dy}{\mu^{2}y^{2}+2(\nu^{2}+\alpha)y+\nu^{2}},
\end{equation}
where
\begin{equation}\label{eq:mu}
	\mu^{2} = 1+2r\nu+\nu^{2}.
\end{equation}

Using a table of integrals \cite{Gradshtein.2015}, we calculate Eq.~(\ref{eq:PF0}) and explicitly express $\Phi (x)$.  There are three different cases:\newline
(i) for $|r|<1$,
\begin{eqnarray}
	\label{eq:PF1}
	\Phi (x)&=& \gamma^{-2}\biggl\{x^{2} +\left(\frac{1}{\beta} -\frac{2\beta}{\mu ^{2} } \right)\arctan \left(\frac{\mu ^{2} x^{2} +\nu ^{2} +\alpha }{\beta} \right) \nonumber\\
	&&+\frac{1+\alpha}{\mu ^{2} }\ln\left[\mu ^{2} x^{4} +2(\nu ^{2} +\alpha)x^{2} +\nu ^{2}\right]\biggr\},
\end{eqnarray}
where
\begin{equation}\label{eq:gamma}
	\gamma^{2} = (\sigma_{1}\mu )^{2},
\end{equation}
(ii) for $r=\pm 1$ ($\alpha\ne -1$),
\begin{equation}
	\label{eq:PF2}
	\Phi (x)=\gamma^{-2} \left[x^{2} -\frac{1}{\mu (\mu x^{2} +\alpha )} +\frac{2}{\mu} \ln\left|\mu x^{2} +\alpha \right|\right],
\end{equation}
where $\mu = 1+\alpha$,\newline
(iii) for $r=-1$ and $\nu =1$ ($\alpha =-1$),
\begin{equation}
	\label{eq:PF3}
	\Phi (x)=\frac{1}{3\sigma_{1}^{2}}\left(1+x^{2} \right)^{3}.
\end{equation}

Substituting Eqs.~(\ref{eq:B(x)}), (\ref{eq:PF1}), (\ref{eq:PF2}), and (\ref{eq:PF3}) into Eq.~(\ref{eq:SPD0}), we respectively obtain the explicit expressions for the SPD of $x(t)$:\newline
(i) for $|r|<1$,
\begin{eqnarray}
	\label{eq:SPD1}
	\mathcal{P}(x)&=&\frac{\mathcal{N}(1+x^{2})}{[\mu ^{2}x^{4}+2(\nu ^{2} +\alpha)x^{2} +\nu ^{2}]^{\frac{1}{2} +\frac{1+\alpha}{(\mu\gamma)^{2}} }}  \nonumber\\
	&&\times\exp \biggl\{-\frac{1}{\gamma^{2}} \biggl[x^{2}
	+\left(\frac{1}{\beta } -\frac{2\beta }{\mu ^{2} } \right)\nonumber\\
	&&\times\arctan \left(\frac{\mu ^{2} x^{2} +\nu ^{2} +\alpha }{\beta } \right)\biggr]\biggr\},
\end{eqnarray}
(ii) for $r=\pm 1$ ($\alpha\ne -1$),
\begin{equation}
	\label{eq:SPD2}
	\mathcal{P}(x)=\frac{\mathcal{N}(1+x^{2})}{\left|\mu x^{2} +\alpha \right|^{1+  \frac{2}{\mu\gamma^{2}} }} \exp\left\{-\frac{1}{\gamma^{2}} \left[x^{2} -\frac{1}{\mu (\mu x^{2} +\alpha)} \right]\right\},
\end{equation}
(iii) for $r=-1$ and $\nu =1$ ($\alpha =-1$),
\begin{equation}
	\label{eq:SPD3}
	\mathcal{P}(x)=\mathcal{N}(1+x^{2})\exp \left[-\frac{1}{3\sigma_{1}^{2}} \left(1+x^{2} \right)^{3} \right].
\end{equation}

The SPDs~(\ref{eq:SPD1}) and (\ref{eq:SPD2}) are defined by three independent parameters $(\sigma_{1}, \sigma_{2}, r)$ that are present in Eqs.~(\ref{eq:SDE1}) and (\ref{eq:NCF}). The other parameters are expressed through them. However, alternative representations are possible. The considered distributions can be written, for instance, in terms of $(\sigma_{1}, \alpha, \beta)$. In this case, $\mu^{2} = (1+\alpha)^{2}+\beta^{2}$ and $\nu^{2} = \alpha^{2}+\beta^{2}$.  The corresponding Langevin equation reads
\begin{equation}
	\label{eq:SDE2}
	\dot{x} = -x + \sigma_{1}G(x)\xi(t),
\end{equation}
where $\xi(t)$ is Gaussian white noise, $\langle\xi(t)\rangle=0$ and $\langle\xi(t)\xi(t')\rangle=\delta(t-t')$, and the multiplicative factor $G(x)$ takes the form
\begin{equation}\label{eq:G(x)}
	G(x)=\left\{ [x^{2} (1+x^{2} )^{-1} +\alpha ]^{2} +\beta ^{2} \right\}^{1/2}.
\end{equation}
Here we convert the two-noise Langevin equation~(\ref{eq:SDE1}) to the statistically equivalent one-noise Langevin equation~(\ref{eq:SDE2}).

The SPD~(\ref{eq:SPD1}) is unbounded and has Gaussian tails, i.e., $\mathcal{P}(x)\propto\exp(-x^{2}/\gamma^{2})$ as $|x|\gg 1$. Indeed, for large $|x|$, the system (\ref{eq:SDE1}) is driven by two additive $\xi_{1}(t)$ and $\xi_{2}(t)$ noises. The quantity $\gamma^{2}/2$ is the intensity of the stochastic process $\sigma_{1}\xi_{1}(t) + \sigma_{2}\xi_{2}(t)$. For small $|x|$, Eq.~(\ref{eq:SPD1}) can be reduced to the asymptotic form
\begin{eqnarray}\label{eq:AF1}
	\mathcal{P}(x)&\propto&\bigg[ 1+\left( \frac{x^{2}+\alpha}{\beta}\right)^{2} \bigg]^{-1/2}\exp\bigg[ -\frac{1}{\beta\sigma_{1}^{2}}\nonumber\\
	 &&\times\arctan\left( \frac{x^{2}+\alpha}{\beta}\right) \bigg],\quad  |x|\ll 1,
\end{eqnarray}
which does not contain the parameter $\gamma$ and is completely determined by the parameters $\alpha$, $\beta$, and $\sigma_{1}$. The dynamics of the system (\ref{eq:SDE1}) is non-Gaussian in the neighborhood of the point $x=0$. We do not find the probability density (\ref{eq:AF1}) in a table of distributions, but it can formally be obtained from the unnormalized probability density of the Pearson type IV distribution \cite{Jeffreys.1961} by substituting $x^2$ instead of $x$.

The SPD~(\ref{eq:SPD2}), if $r=1$ or $r=-1$ and $\sigma_{2}>\sigma_{1}$, is also unrestricted with Gaussian tails. But, if $r=-1$ and $\sigma_{2}<\sigma_{1}$, it is restricted to the interval (\ref{eq:domen)}). For small $|x|$, the asymptotics of Eq.~(\ref{eq:SPD2}) can be written as
\begin{equation}\label{eq:AF2}
	\mathcal{P}(x)\propto \frac{1}{|x^{2}+\alpha|}\exp\left[ \frac{1}{\sigma_{1}^{2}(x^{2}+\alpha)}\right], \quad |x|\ll 1.
\end{equation}
If $r=-1$ and $\sigma_{2}=\sigma_{1}$, Eq.~(\ref{eq:AF2}) is the unnormalized SPD of the genetic model \cite{Horsthemke.1984,Arnold.1978,Mahnke.2009}. So, the considered model (\ref{eq:SDE1}) is close to it in the specific case.

The SPD~(\ref{eq:SPD3}) ($r=-1$ and $\sigma_{2}=\sigma_{1}$) is unrestricted, but its tails are non-Gaussian, $\mathcal{P}(x)\propto x^{2}\exp(-x^{6}/3\sigma_{1}^{2})$ as $|x|\gg 1$. This asymptotics agrees with that obtained for the system driven by Gaussian white noise, whose amplitude depends on the state variable $x$ as $x^{-2}$ \cite{Denisov.2002}. For large $|x|$, the additive noises $\xi_{1}(t)$ and $\xi_{2}(t)$ exactly cancel each other, $\gamma^{2}=0$, and the system's dynamics is deterministic. The Langevin equation (\ref{eq:SDE2}) related to Eq.~(\ref{eq:SPD3}) takes the form
\begin{equation}
\label{eq:SDE3}
\dot{x} = -x + \sigma_{1}(1+x^{2})^{-1} \xi_{1}(t).
\end{equation}
The multiplicative factor $G(x)=1/(1+x^{2})$ coincides with the field-dependent kinetic coefficient for the spatially extended systems \cite{Ibanes.2001, Buceta.2003} in which the zero-dimensional units demonstrate noise-induced transitions. For $|x|\gg 1$, $G(x)$ vanishes and the noise $\xi(t)$ does not affect the systems. For $|x|\ll 1$, $G(x)\sim 1-x^{2}$, Eq.~(\ref{eq:SDE3}) is the Langevin equation of the genetic model.

\section{Phase diagram and order parameter}\label{sec:PhD_OP}

\begin{figure}
	\includegraphics{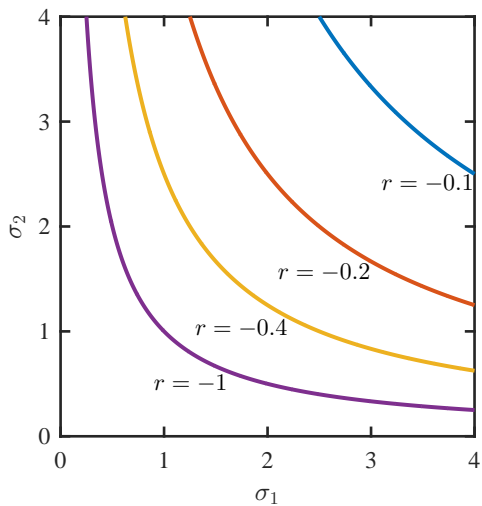}
	\caption{\label{fig1} Phase diagram in the $(\sigma_{2}, \sigma_{1})$ plane for Eq.~(\ref{eq:D}). The region above the corresponding critical curve is bimodal.}
\end{figure}

\begin{figure*}
	\includegraphics{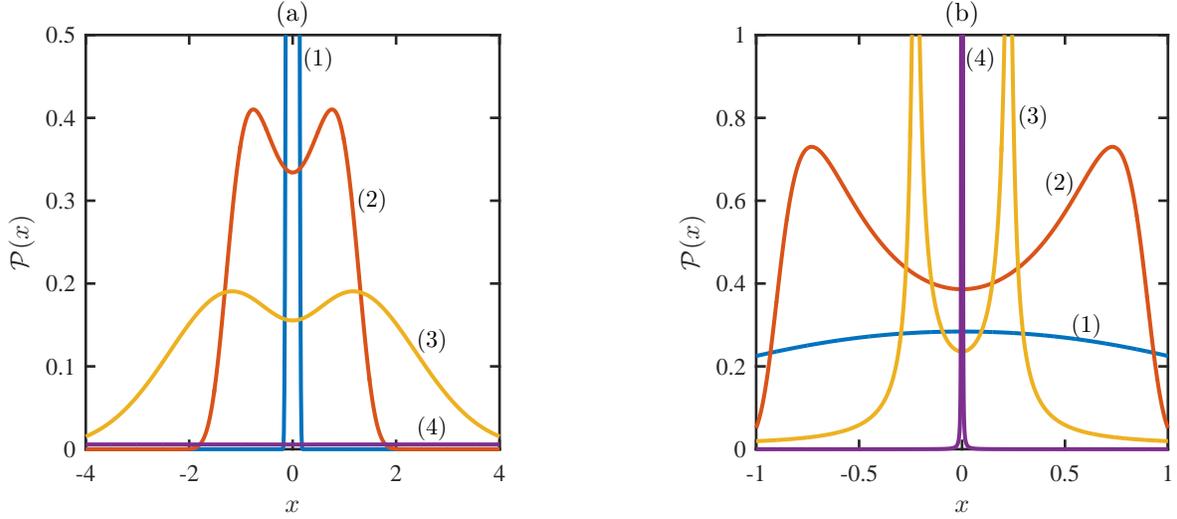}
	\caption{\label{fig2} Plot of the SPD $\mathcal{P}(x)$ vs the state variable $x$ for Eq.~(\ref{eq:SPD1}), $r=-0.99$. (a)~The amplitude of the multiplicative noise is fixed, $\sigma_{1}=2$. The amplitude of the additive noise increases: (1)~$\sigma_{2}=0.1$, (2)~$\sigma_{2}=2$, (3)~$\sigma_{2}=4$, (4)~$\sigma_{2}=10^{2}$. The curve (1) is unimodal. The curve (4) is bimodal, but its two maxima are indistinguishable and spaced a considerable distance apart. (b)~The amplitude of the additive noise is fixed, $\sigma_{2}=2$. The amplitude of the multiplicative noise increases: (1)~$\sigma_{1}=0.1$, (2)~$\sigma_{1}=4$, (3)~$\sigma_{1}=40$, (4)~$\sigma_{1}=10^{5}$. The curve (4) is bimodal, but its two maxima are close enough to each other.}   
\end{figure*}

According to Ref.~\cite{Denisov.2003}, the shape of the SPDs~(\ref{eq:SPD1}), (\ref{eq:SPD2}), and (\ref{eq:SPD3}) is characterized by the following parameter:
\begin{equation}\label{eq:D}
	D=r\sigma_{1}\sigma_{2}
\end{equation}
with the critical value $D_{c}=-1$. Eq.~(\ref{eq:D}) is used to plot the phase diagram of the steady-state behavior of the model (\ref{eq:SDE1}) in the $(\sigma_{2}, \sigma_{1})$ plane for different values of $r$ (FIG.~\ref{fig1}). The extreme points of the SPD are the order parameter for transitions. If $D>D_{c}$, the SPD of $x(t)$ is bell-shaped with a single maximum at $x=0$, the preferential state coincides with the deterministic steady state. If $D<D_{c}$, $\mathcal{P}(x)$ has a crater-like form with two maxima at $x_{m\pm}$ and a local minimum at $x=0$, two preferential states arise, whereas the underlying deterministic dynamics has only one stable state. Hence at $D=D_{c}$, the SPD undergoes a transition from unimodal to bimodal distribution. It is possible only if the anti-correlation exists, $-1\leqslant r<0$, as indicated by the minus sign in $D_{c}$. In this case, by increasing both the amplitude $\sigma_{1}$ of the multiplicative noise and the amplitude $\sigma_{2}$ of the additive noise, bistable states are created (FIG.~\ref{fig2}). Thus, both additive and multiplicative noises can induce transitions in the presence of negative cross-correlation. But if they are not sufficiently strong, $\sigma_{1}<1$ and $\sigma_{2}<1$, or if one noise is strong and the other one is weak that $\sigma_{1}\sigma_{2}<1$, by changing the strength $r$ of the correlation between the noises, the SPD remains unimodal. In these cases, the cross-correlation can not lead to transitions. Therefore, we say that bistability is induced by two cross-correlated Gaussian white noises.

The maximum points $x_{m\pm}$ of the bimodal SPDs~(\ref{eq:SPD1}), (\ref{eq:SPD2}) and (\ref{eq:SPD3}) are found from Eq.~\cite{comment.1}. It yields
\begin{equation}\label{eq:cubic}
	y^{3}+3py+2q=0\quad (y\geqslant 1),
\end{equation}
where $3p=r\sigma_{1}\sigma_{2}+\sigma_{1}^{2}$ and $2q=-\sigma_{1}^{2}$. If $D\leqslant D_{c}$, the solution of the cubic equation~(\ref{eq:cubic}) can be written as follows (see, for instance, Ref.~\cite{Bronshtein.2015}):\newline
(i) for $p<0$ and $q^{2}+p^{3}\leqslant 0$,
\begin{equation}\label{eq:y1}
	y=2\sqrt{|p|}\cos\left[ \frac{1}{3}\arccos\left( -\frac{q}{\sqrt{|p|^{3}}}\right) \right], 
\end{equation}
(ii) for $p<0$ and $q^{2}+p^{3}>0$,
\begin{equation}\label{eq:y2}
y=2\sqrt{|p|}\cosh\left[ \frac{1}{3} \mbox{arcosh}\left( -\frac{q}{\sqrt{|p|^{3}}}\right) \right], 
\end{equation}
(iii) for $p>0$,
\begin{equation}\label{eq:y3}
y=2\sqrt{p}\sinh\left[ \frac{1}{3} \mbox{arsinh}\left( -\frac{q}{\sqrt{p^{3}}}\right) \right],
\end{equation}
(iv) for $p=0$ ($r=-1$ and $\sigma_{1}=\sigma_{2}$),
\begin{equation}\label{eq:y4}
	y=\sigma_{1}^{2/3},
\end{equation}
and
\begin{equation}\label{eq:xm}
	x_{m\pm}=\pm\sqrt{y-1}.
\end{equation}

The role that the noises play in the considered phenomenon is not just creating bistability.

\begin{figure*}
	\includegraphics{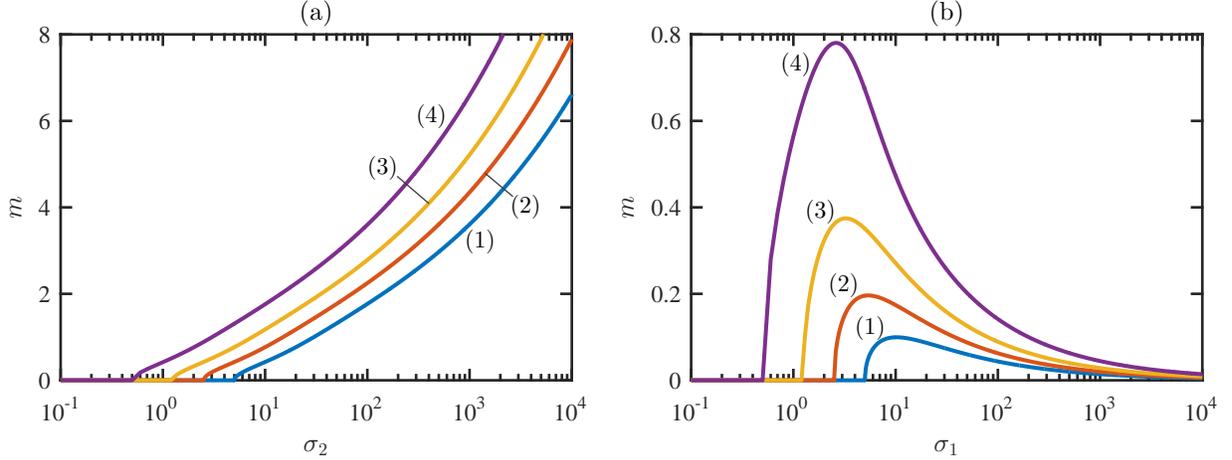}
	\caption{\label{fig3} Order parameter $m=|x_{m\pm}|$ vs (a)~the amplitude $\sigma_{2}$ of the additive noise or (b)~the amplitude $\sigma_{1}$ of the multiplicative noise for Eq.~(\ref{eq:xm}). The coefficient $r$ of the correlation between the noises is varied: (1)~$r=-0.1$, (2)~$r=-0.2$, (3)~$r=-0.4$, (4)~$r=-0.99$. (a)~The parameter $\sigma_{1}$ is fixed, $\sigma_{1}=2$. The critical value is $\sigma_{2c}=-1/r\sigma_{1}$. If $\sigma_{2}<\sigma_{2c}$, $m=0$.  If $\sigma_{2}>\sigma_{2c}$, $m$ increases monotonically with $\sigma_{2}$. (b)~The parameter $\sigma_{2}$ is fixed, $\sigma_{2}=2$. The critical value is $\sigma_{1c}=-1/r\sigma_{2}$. If~$\sigma_{1}<\sigma_{1c}$, $m=0$. If $\sigma_{1}>\sigma_{1c}$, $m$ first increases to a maximum value, then decreases with $\sigma_{1}$, and $m\rightarrow 0$ as $\sigma_{1}\rightarrow\infty$.}   
\end{figure*}

The additive noise has also a disordering effect, it spreads the SPD as illustrated in Fig.~\ref{fig2}(a). The order parameter $m=|x_{m\pm}|$ is equal to zero if $\sigma_{2}<\sigma_{2c}$, where $\sigma_{2c}=-1/r\sigma_{1}$, and increases monotonically with increasing $\sigma_{2}$ if $\sigma_{2}>\sigma_{2c}$ [see Fig.~\ref{fig3}(a)]. Taking into account Eqs.~(\ref{eq:y1}) and (\ref{eq:xm}), we have
\begin{equation}\label{eq:m1}
	m\sim\sqrt{\sqrt{-r\sigma_{1}\sigma_{2}}-1}\quad (\sigma_{2}\rightarrow\infty).
\end{equation}
The distance $2m$ between two maxima of $\mathcal{P}(x)$ tends to infinity as the intensity of the additive noise increases indefinitely. The additive noise, correlated with the multiplicative noise, induces bistable states firstly, then destroys them in the end. But formally the SPD remains bimodal.

In contrast to the additive noise, the multiplicative noise has an \textit{ordering} effect: by increasing its intensity, the SPD is narrowed [see Fig.~\ref{fig2}(b)] and two sharp peaks appear [the curve (3)]. The order parameter $m$ is zero if $\sigma_{1}<\sigma_{1c}$, where $\sigma_{1c}=-1/r\sigma_{2}$, and first increases to a maximum value, then starts to decrease with $\sigma_{1}$ if $\sigma_{1}>\sigma_{1c}$ [see Fig.~\ref{fig3}(b)]. According to Eqs.~(\ref{eq:y3}) and (\ref{eq:xm}), the asymptotics of $m$ for large $\sigma_{1}$ takes the form
\begin{equation}\label{eq:m2}
	m\sim\sqrt{\frac{1}{1+r\nu}-1}\quad (\nu\rightarrow 0),
\end{equation}
it reaches zero as $\sigma_{1}\rightarrow\infty$. Two sharp peaks approach each other and converge into a single peak around $x=0$. In this case, the SPD is still bimodal, its two maxima coincide. It is important that the phase diagram in FIG.~\ref{fig1} does not show any reentrant transitions. The multiplicative noise, correlated with additive noise, induces bistability with a critical point at $x=0$ and $\sigma_{1}=\sigma_{1c}$ where the SPD has a double maximum. In fact, a \textit{second} critical point occurs at $x=0$ and $\sigma_{1}\rightarrow\infty$ that is an unexpected result. It can be interpreted by analyzing Eqs.~(\ref{eq:SDE2}) and (\ref{eq:G(x)}). The parameters~(\ref{eq:alpha_beta)}) $\alpha$ and $\beta$ tend to zero as $\sigma_{1}\gg\sigma_{2}$. As a result, the Langevin equation~(\ref{eq:SDE2}) is reduced to the Langevin equation~(\ref{eq:SDE1}) with $\xi_{2}(t)\rightarrow 0$. In this case, as mentioned in Sec.~\ref{sec:model}, the point $x=0$ is absorbing and the SPD is the Dirac delta function.

An ordering effect of the multiplicative noise is unusual. Obviously, we expect a disordering effect from any external noise \cite{Horsthemke.1984}, as in our case $r=-1$ and $\sigma_{2}=\sigma_{1}$ [see Eqs.~(\ref{eq:SDE3}), (\ref{eq:y4}), and (\ref{eq:xm})]. The order parameter
\begin{equation}\label{eq:m3}
	m=\sqrt{\sigma_{1}^{2/3} - 1}
\end{equation}
diverges as $\sigma_{1}\rightarrow\infty$ and the SPD~(\ref{eq:SPD3}) is broadened. It should be noted that the multiplicative factor $G(x)=1/(1+x^{2})$ does not vanish at $x=0$ and the absorbing boundary does not appear.

It is known that noise-induced bistability is the competitive effect between a deterministic restoring force and a multiplicative random force. The former drives the system toward the steady state, $x=x_{0}$, the latter away from it. The multiplicative factor must have a maximum at $x=x_{0}$ and decrease as $x$ deviates from $x_{0}$ \cite[pp.~219-220]{Ridolfi.2011}.  More concretely, if the deterministic force is linear, $f(x)=-x+o(x)$, then the multiplicative factor must be in the form $g(x)=1-x^{2}+o(x^{2})$ \cite{VandenBroeck.1997}. When the noise is sufficiently strong, the bimodal SPD of $x$ may occur as a result of the balance between deterministic and random forces. In order to understand the mechanism of the phenomenon of bistability induced by two cross-correlated Gaussian white noises, we analyze the one-noise Langevin equation (\ref{eq:SDE2}). It has two key ingredients: (1)~the linear restoring force, $f(x)=-x$; (2)~the multiplicative noise term, $\sigma_{1}G(x)\xi(t)$. The latter is the result of the cooperative interaction of two cross-correlated multiplicative and additive Gaussian white noises~(\ref{eq:SDE1}). The Maclaurin series expansion of $G(x)$~(\ref{eq:G(x)}) is given by
\begin{equation}\label{eq:MS}
G(x)=\nu+rx^{2}+o(x^{4}).
\end{equation}
Eq.~(\ref{eq:MS}) indicates that if the cross-correlation is negative, the multiplicative factor $G(x)$ has appropriate form for the emergence of noise-induced transitions.

\section{Conclusions}\label{sec:Concl}

The prototype model of a stochastic zero-dimensional one-variable system with a linear restoring force driven by two cross-correlated multiplicative and additive Gaussian white noises has been studied in Ref.~\cite{Denisov.2003}. The multiplicative factor has been assumed to be quadratic for small absolute values of the state variable and constant for large ones. It has been shown that by changing the strength of a negative cross-correlation the SPD of the state variable may undergo a transition from unimodal to bimodal distribution. In this paper, we have investigated this model in more detail. We have calculated explicitly the SPD of the state variable and found its asymptotics. It has been determined that if the noises are perfectly anti-correlated ($r=-1$) with equal intensities, the considered model is reduced to the well-known genetic model for small absolute values of the state variable. We have constructed the phase diagram and shown that by increasing both the intensity of the multiplicative noise and the intensity of the additive noise, bistability may emerge when a negative cross-correlation is present. We have obtained the order parameter for transitions corresponding to the extreme points of the SPD. We determined that the additive noise has a disordering effect on the stationary behavior of the system, it spreads the SPD and tends to destroy the deterministic steady state and bistable states. The multiplicative noise has an ordering effect, it narrows the SPD and tends to stabilize the deterministic steady state, and a second critical point occurs. It has been specified the key ingredients of bistability. They are: (1) linear restoring force; (2) additive Gaussian white noise; (3) multiplicative Gaussian white noise, whose amplitude depends on the state variable $x$ as $x^{2}$ in the vicinity of the steady state $x=0$; (4) negative cross-correlation. Anti-correlation between the noises means that if one noise increases, the other decreases, and vice versa. Their cooperative interplay causes the effective stochastic force, whose amplitude takes a maximum value at $x=0$ and decreases as $|x|$ increases. The deterministic restoring force drives the system toward the steady state, the effective stochastic force away from it, and bistability may occur as a result of the balance between forces. Although the model is abstract, we expect that the same phenomenon can be observed in real systems.

\bibliography{mypaper}

\begin{thebibliography}{48}
\bibitem [{com()}]{comment.1}%
  \BibitemOpen
  \href@noop {} {}\bibinfo {note} {The extrema of the SPD for the system with
  the force $f(x)=-x$ subject to cross-correlated multiplicative
  $g(x)\xi_{1}(t)$ and additive $\xi_{2}(t)$ noises are found from
  \cite{Denisov.2003} \[ \sigma_{1}^{2}gg'+r\sigma_{1}\sigma_{2}g' + 2x = 0. \]
  By analyzing this equation, one can conclude that $g(x)=x^{2}$ is the
  simplest case in which the number of extrema changes from 1 (one maximum) to
  3 (two maximum and one minimum) as the coefficient of the cross-correlation
  $r$ is varied. Moreover, if $r=0$ no changes are possible.}\BibitemShut
  {Stop}%
\end{thebibliography}

\end{document}